# Modelling Reactive and Proactive Behaviour in Simulation: A Case Study in a University Organisation

Mazlina Abdul Majid [1], Peer-Olaf Siebers[2], Uwe Aickelin [2],

[1]Universiti Malaysia Pahang, Malaysia
mazlina@ump.edu.my

[2]University of Nottingham, UK
{pos, uxa}@cs.nott.ac.uk

**Abstract:** Simulation is a well established what-if scenario analysis tool in Operational Research (OR). While traditionally Discrete Event Simulation (DES) and System Dynamics Simulation (SDS) are the predominant simulation techniques in OR, a new simulation technique, namely Agent-Based Simulation (ABS), has emerged and is gaining more attention. In our research we focus on discrete simulation methods (i.e. DES and ABS). The contribution made by this paper is the comparison of DES and combined DES/ABS for modelling human reactive and different level of detail of human proactive behaviour in service systems. The results of our experiments show that the level of proactiveness considered in the model has a big impact on the simulation output. However, there is not a big difference between the results from the DES and the combined DES/ABS simulation models. Therefore, for service systems of the type we investigated we would suggest to use DES as the preferred analysis tool.

## 1 Introduction

Simulation has become a preferred tool in Operation Research for modelling complex systems (Kelton et al. 2007). Simulation is considered a decision support tool which has provided solutions to problems in industry since the early 1960s (Shannon 1975). Studies in human behaviour modelling and simulation have received increased focus and attention from simulation research in the UK (Robinson, 2004). Human behaviour modelling and simulation refers to computer-based models that imitate either the behaviour of a single human or the collective actions of a team of humans (Pew and Mavor 1998).

Discrete Event Simulation (DES) and Agent Based Simulation (ABS) are simulation approaches often used for modelling human behaviour in Operational Research (OR). Examples can be found in Brailsford et al. (2006) and Siebers et al (2010). The capability of modelling human behaviour in both simulation approaches is due to their ability to model heterogeneous entities with individual behaviour. Another simulation approach commonly used in OR is System Dynamic (SD). However, this approach focuses on modelling at an aggregate level and is therefore not well suited to model a heterogeneous population at an individual level. Because of this limitation, SDS is not considered in the present study.

Human behaviour can be categorised into different types, many of which can be found in the service sector. When talking about different kinds of human behaviour we refer to reactive and proactive behaviour. Here, reactive behaviour is related to staff responses to the customer when being requested and available while proactive behaviour relates to a staff member's personal initiative to identify and solve an issue. When providing services proactivity of staff plays an important role in an organisation's ability to generate income and revenue (Rank et al, 2007).

But the question here- *is it useful to consider proactive behaviour in models of service systems and which simulation techniques is the best choice for modelling such behaviour?* Thus, in this paper we investigate the impact of modelling different levels of proactive behaviour in DES and combined DES/ABS. We compare both simulation techniques in term of simulation result using a real world case study: The International Office at the University of Nottingham (University of Nottingham, 2011).

Previously we have studied the capabilities of DES and combined DES/ABS in representing the impact of reactive staff behaviour (Majid et al, 2009) and mixed reactive and proactive behaviour (Majid et al, 2010) in a retail sector environment. In this paper we look at the capabilities of DES and combined DES/ABS in representing the impact of mixed reactive and proactive behaviour on public sector systems.

The paper is structured as follows: In Section 2 we explore the characteristics of DES and ABS and discuss the existing literature on modelling human behaviour in service sector. In Section 3 we describe our case study and the simulation models development and implementation. In Section 4 we present our experimental setup, the results of our experments, and a discussion of these results. Finally, in Section 5 we draw some conclusions and summarise our current progress.

## 2 Literature Review

### 2.1 Human Behaviour Modelling in DES and ABS

This chapter reviews existing research studies on modelling human behaviour using DES and ABS techniques. As explained by Pew and Mavor (1998), Human Behaviour Representation (HBR), also known as human behaviour modelling, refers to computer-based models which imitate either the behaviour of a single person or the collective actions of a team of people. Nowadays, research into human behaviour modelling is well documented globally and discussed in a variety of application areas. Simulation appears to be the preferred choice as a modelling and simulating tool for investigating human behaviour (ProModel 2010). This is because the diversity of human behaviours is more accurately depicted by the use of simulation (ProModel 2010).

Throughout the literature, the best-known simulation techniques for modelling and simulating human behaviour are DES and ABS. Among existing studies on modelling human behaviour, the use of DES is presented by Brailsford et al. (2006), Nehme et al. (2008) and Baysan et al.(2009). On the other hand, Schenk et al. (2007), Siebers et al. (2007) and Korhonen et al. (2008a; 2008b) recommend ABS for modelling human behaviour.

Brailsford et al. (2006) claim that, based on their experiments of modelling the emergency evacuation of a public building, it is possible to model human movement patterns in DES. However, the complex nature of DES structures where entities in the DES model are not independent and self-directed makes the DES model inappropriate for modelling large-scale systems. This characteristic of entities in DES is agreed by Baysan et al.(2009), who have used DES in planning the pedestrian movements of the visitor to the Istanbul Technical University Science Center. However, due to the dependent entities in the DES model, the pedestrian movement pattern in their simulation model is restricted to pre-determined routes. By contrast, Korhonen (2008a; 2008b) has developed an agent-based fire evacuation model which models people-flow in free movement patterns. He states that the decision to use ABS is due to the fact that agent-based models can provide a realistic representation of the human body with the help of autonomous agents. In addition to modelling human behaviour using DES, Nehme et al (2008) have investigated methods of estimating the impact of imperfect situational awareness of military vehicle operators. They claim that it is possible to use the DES model to understand human behaviour by matching the results from the DES model with human subjects. Schenk et al. (2007) comment that modelling consumer behaviour when grocery shopping is easier using ABS because this model has the ability to integrate communication among individuals or consumers. Siebers et al. (2007) assert that their research in applying ABS to simulate management practices in a department store appears to be the first research study of its kind. They argue that ABS is more suitable than DES due to the characteristics of the ABS model; specifically, it allows to model proactive and autonomous entities that can behave similar to humans in a real world system.

Instead of choosing only one simulation technique to model human behaviour, some researchers tend to combine DES and ABS in order to model a system which cannot be modelled by either method independently. Examples are Page et al. (1999) who studied the operation of courier services in logistics, Kadar et al. (2005) who investigated manufacturing systems, Dubiel and Tsimhoni (2005) who studied human travel systems and Robinson (2010) who looked at the operation of coffee shop services. They all agree that DES and ABS modelling can complement each other in achieving their objectives. A combination of ABS and DES modelling is useful when human behaviour has to be modelled for representing communication and autonomous decision-making.

In conclusion, the research into human behaviour using DES and ABS that has been carried out so far suggests that DES and ABS are able to model human behaviour but take different approaches (dependent entities vs. independent agents). The studies outlined above indicate that DES is suitable for capturing simple human behaviour, but is problematic when applied to more complex behaviours as the next event to occur in DES has to be determined. In contrast, ABS offers straightforward solutions to modelling complex human behaviour, i.e. free movement patterns or employee proactive behaviour, as agents can initiate an event themselves. However, for ABS the resource requirements (computational power) are much higher and the modelling and implementation of the model is more complex.

**2.2 The Simulation Choice**

For our comparison exercise we have chosen to use DES and combined DES/ABS. As we are interested to investigate a service oriented system in the public sector, which involves queuing for the different services we cannot use pure ABS for our investigations, as in pure ABS models the system itself is not explicitly modelled but emerges from the interaction of the many individual entities that make up the system. However, as ABS seems to be a good concept for representing human behaviour we use a combined DES/ABS approach where we model the system in a process-oriented manner while we models the actors inside the system (the people) as agents.

# 3 Case Study Description

The subject of our case study is the delivery of international support services at the University of Nottingham in the UK (University of Nottingham, 2011) which is one of the world's most prominent universities. One of the main reasons for this choice is that there is frequent interaction between students and support services staff, and interaction behaviour is an important part of the present study into human behaviour.

The International Support Services Team (ISST) is located in the International Office, offering a wide range of support for the International and European Union students; the office is open from 9.00 am to 5.00 pm every weekday. The present research focuses on the international support operation based at the reception area and within the advisory service. To gain insight into the ISST operation we have conducted observations and collected some data for a period of a week. Figure 1 illustrates the delivery process of support services by the ISST in the University's International Office, the numbers and red arrows representing the sequence of operation.

In ISST, there is one member of staff (receptionist) who works at the reception area and two staff for the advisory service (advisories) who give support over the phone and also offer a one-to-one support service. First, the arriving students or incoming phone calls are served by the receptionist (Figure 1, step 1). At the reception area, the receptionist has to deliver two types of service support tasks: (1) serve the incoming students at the reception desk and (2) respond to incoming phone calls. General enquiries and support requests from a student (i.e. enquiry on a visa presentation schedule) made either in person or on the phone are handled by the receptionist, whose support is available for the whole day. At some point students leave the reception area or the phone calls end after being served by the receptionist (Figure 1, step 2 right).

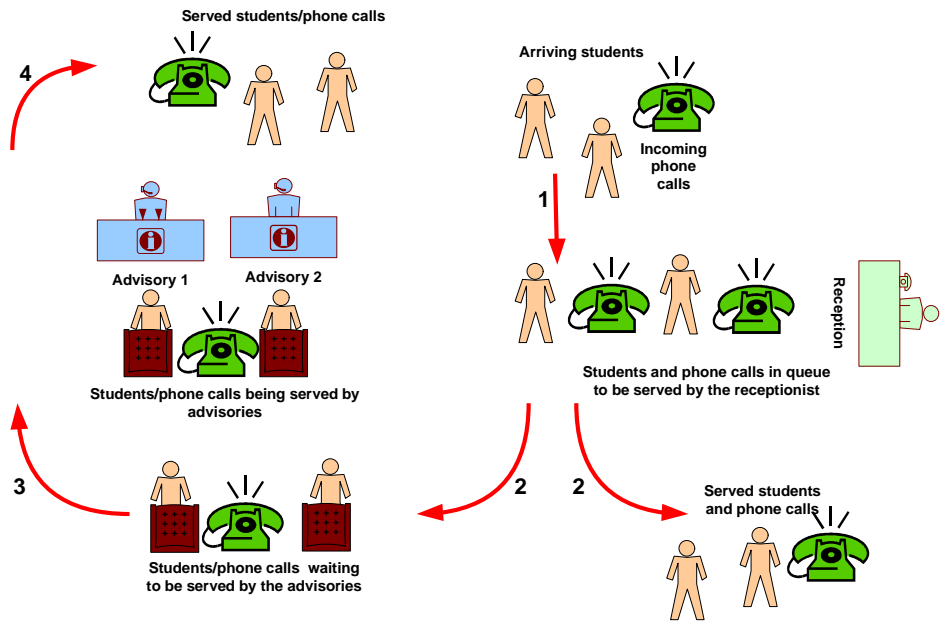

Figure 1: Delivery of the support services in the ISST at the University's International Office

Students can also get help from the advisory service via the walk-in section which is accessible from 1.00 pm to 4.00 pm. A student who wishes to meet with the advisor in the afternoon is required to complete a request form at the receptionist area. The receptionist then gives the student a waiting number and the advisor calls the student when it is his/her turn (Figure 1, step 2 left). The number and the form that is completed earlier are collected by the advisor on duty before serving the students (Figure 1, step 3). The student leaves the advisory section after obtaining the required support (Figure 1, step 4).

During our case study observations we have identified four reactive reception staff behaviours: (1) accepting requests from students in person or on the phone, (2) providing general support to students face to face and during incoming calls, (3) searching for information, and (4) giving waiting numbers to students. Furthermore, we have identified one reactive advisor staff behaviour: to provide detailed support to students in person. Finally, we have identified one reactive student behaviour: to wait in the queue if the staffs (receptionist/advisors) are no available. This relates to students either being there in person or phoning in.

With regard to proactive behaviour, on the other hand, the receptionist is observed to cease handing out waiting numbers if, in their view, there are too many students waiting in the remaining time to be served by the advisors. Their decision to stop handing out waiting numbers is based on monitoring experience at the ISST operation. The advisors demonstrate proactive behaviour in speeding up their service time to ensure that all students who are waiting are served in the remaining operation time, a decision that is also based on the experience in serving students. The proactive behaviour observed in the students is skipping the queue in order to ask the receptionist a question. The decision to skip the queue is initiated from observing the queue at reception.

## 4 Simulation Models

Both, the DES model and the combined DES/ABS model are based on the same conceptual model (Figure 1) but the strategy of implementation in both cases is very different. DES modelling uses a process-oriented approach, i.e. the development begins by modelling the basic process flow of the ISST operations as a queuing system (Figure 2a). Then, the investigated human behaviours, reactive and proactive (Figure 2a/b), are added to the basic process flow.

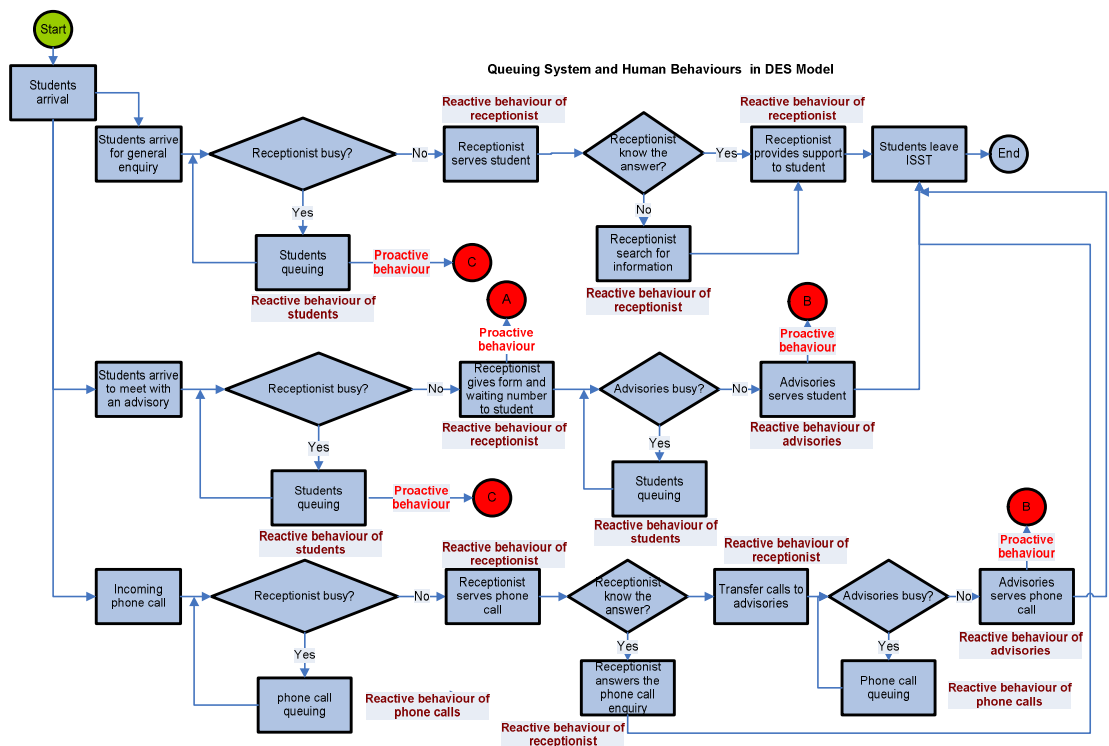

Figure 2a: The implementation of DES model (process flow + reactive behaviours)

Two different implementation approaches are used for developing the combined DES/ABS model: the process-oriented approach is used for the DES modelling (Figure 2a) and the individual-centric approach is used to model the agents. Figure 3 shows some state charts that represent the different types of agents in the model (students/phone calls, receptionist and advisories).

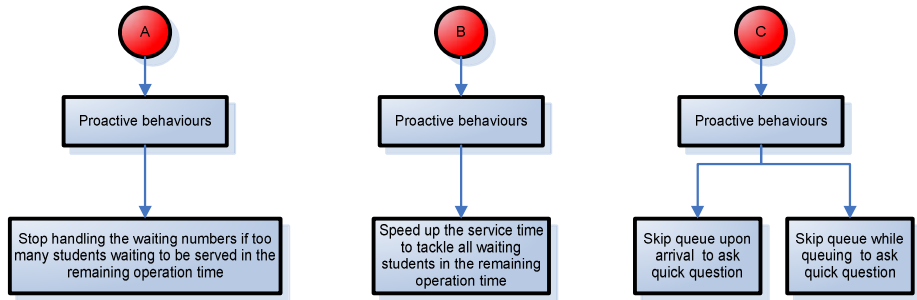

Figure 2b: The implementation of DES model (proactive behaviours)

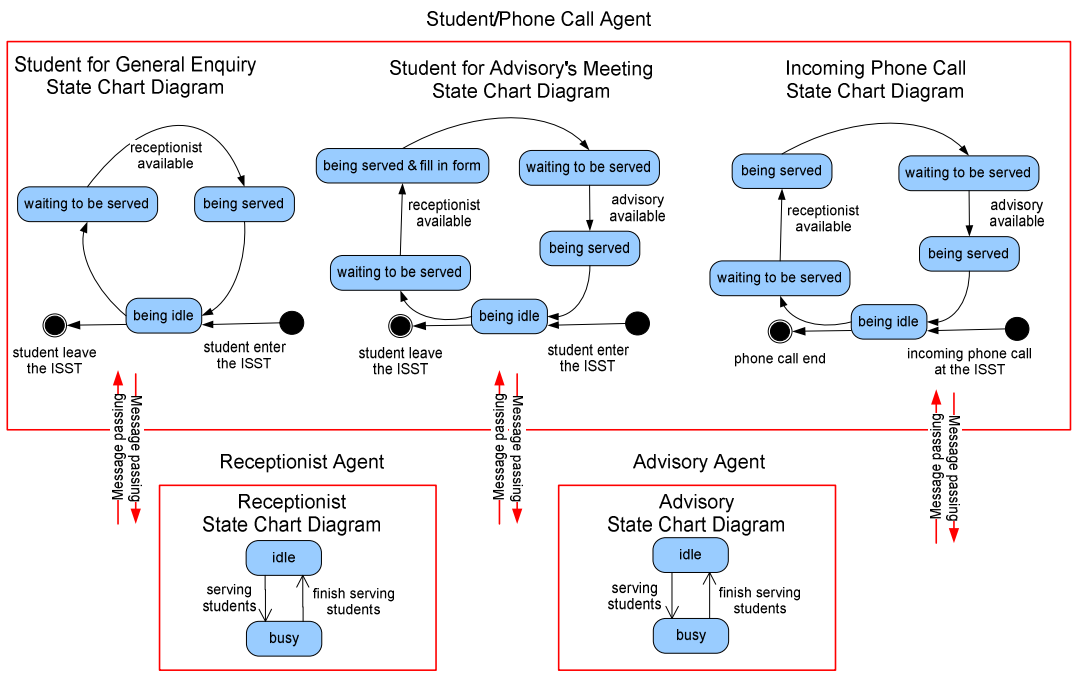

Figure 3: The implementation of combined DES/ABS model

The simulation models have been built in AnyLogic[TM] 6.5, a Java based multi-paradigm simulation software. Both simulation models consist of three arrival processes (students' arrival for general enquiry, students' arrival for advisory meeting, and incoming phone calls), one single queue for each arrival, and three resources (one receptionist and two advisors). In the DES model, student/phone call, receptionist and advisors are all passive objects while in the combined DES/ABS model these are active objects (agents). Passive objects are entities that are affected by the simulation's elements as they move through the system, while active objects are entities acting themselves by initiating actions (Siebers et al., 2010).

The student arrival rate has been modelled using an exponential distribution with an hourly changing rate in accordance with the arrival pattern we observed in the real system. The simulation inputs for receptionist service time and advisors service time are modelled using triangular distributions. The required values for the triangular distributions have been obtained by calculating the daily minimum, average and maximum service times observed during the case study period.

The simulation models terminate after a standard business day (8 hours), mimicking the operation of the real system. We conducted 100 replications for each set of parameters. Both simulation models use the same model input parameter values. Therefore, if we see any differences in the model outputs they will be due to the impact of the differences between the modelling techniques. The verification and validation process are performed simultaneously with the development of the basic simulation models (DES and combined DES/ABS).

## 5 Experiment

The purpose of the experiment is to compare the simulation results when modelling reactive human behaviour against mixed reactive and proactive human behaviour in both DES and combined DES/ABS models. The comparison of simulation results for DES and combined DES/ABS models is conducted statistically by performing a t-test. For our comparison we use the following hypotheses:

$Ho_1$ :     Reactive and mixed reactive and proactive behaviour DES models produce statistically the same simulation results.

$Ho_2$ :     Reactive and mixed reactive and proactive behaviour combined DES/ABS models produce statistically the same simulation results.

Five types of experiments have been conducted as shown in Table 1. In this case study, the general idea of reactive behaviour and a specific type of proactive behaviour (Type 1, Type 2 and a combination of both types) were modelled. Experiment 1 is investigating general reactive behaviour (response to environment). Experiment 2 and 3 are studying Type 1 proactive behaviour. This proactive behaviour is related to the behaviour of receptionist and advisors when making their own decisions, based on their experience, to deal with the hectic situation in the ISST. Two proactive behaviours are investigated: Type 1a - The receptionist stops handing out waiting numbers when there are too many students to be served by the advisors in the remaining time (Experiment 2) and Type 1b - Advisors speed up their service time to ensure that all students waiting to be served are supported in the remaining operation time (Experiment 3). Experiment 4 is looking at Type 2 proactive behaviour. This proactive behaviour refers to the observed behaviour of students in achieving their aim. The students skip queues in order to ask the receptionist a question. Experiment 5 considers the combined behaviour of Experiment 1 to 4.

Table 1 : Experiments

| Behaviours | | Experiment |
|---|---|---|
| Type | Sub-Type | Sub-Experiment |
| General | Reactive : React to students and incoming calls request | Experiment 1 |
| 1 | Sub-Proactive 1 : Customer request to leave | Experiment 2 |
| | Sub-Proactive 2 : Staff speed up service time | Experiment 3 |
| 2 | Sub-Proactive 3 : Customer Skipping from queuing | Experiment 4 |
| 1 and 2 | Sub-Proactive 4 : Combination of Sub-Proactive 1,2 and 3 | Experiment 5 |

Experiment 1 is the reference point as it only contains reactive behaviour. Next we compared the Experiment 1 against the Experiment 2, 3, 4 and 5 for both DES and combined DES/ABS models using "customer waiting time" and "number of customers served" as our performance measures (Table 2). We have selected these measures as the literature recommends them as important measures to increase productivity in the service-oriented systems (Robert and Peter, 2004). It is assumed that investigating these measures will provide sufficient evidence in understanding the impact of the simulation outputs in the different behaviours in one simulation technique.

The sub-hypotheses are built for each performance measure in DES and combined DES/ABS according to the list of experiments to be compared (e.g. $Ho_{1\_D}$ : The waiting time at receptionist resulting from the DES model is not significantly different in Experiments 1 and 2). Finally, the results of the performance measures in the Experiment 1 against Experiment 2, 3, 4 and 5 are gathered and compared for both simulation models (Table 3).

Table 2 : The data of the chosen performance measures for the correlation comparison

| Experiment | DES | | Combined DES/ABS | |
|---|---|---|---|---|
| | Customers waiting time (minutes) | Number of customers not served | Customers waiting time (minutes) | Number of customers not served |
| 1 | 1.43 | 6 | 1.41 | 5 |
| 2 | 1.29 | 0 | 1.25 | 0 |
| 3 | 1.30 | 0 | 1.24 | 0 |
| 4 | 1.24 | 0 | 1.22 | 0 |
| 5 | 1.22 | 0 | 1.21 | 0 |

Table 3: Results for t- test comparing Experiment 1 with 2, 3, 4 and 5

| Experiments | Performance measures | DES P-Value | Ho_DES | Compare with significant level (0.05) | DES/ABS P-Value | Ho_DES/ABS | Compare with significant level (0.05) |
|---|---|---|---|---|---|---|---|
| 1 vs. 2 | Waiting time | 0.0041 | $Ho_{1\_D}$ | Reject | 0.0022 | $Ho_{1\_A}$ | Reject |
| | Customers not served | 0.0000 | $Ho_{2\_D}$ | Reject | 0.0000 | $Ho_{2\_A}$ | Reject |
| 1 vs. 3 | Waiting time | 0.0105 | $Ho_{3\_D}$ | Reject | 0.0141 | $Ho_{3\_A}$ | Reject |
| | Customers not served | 0.0000 | $Ho_{4\_D}$ | Reject | 0.0000 | $Ho_{4\_A}$ | Reject |
| 1 vs. 4 | Waiting time | 0.0033 | $Ho_{5\_D}$ | Reject | 0.0104 | $Ho_{5\_A}$ | Reject |
| | Customers not served | 0.0000 | $Ho_{6\_D}$ | Reject | 0.0000 | $Ho_{6\_A}$ | Reject |
| 1 vs. 5 | Waiting time | 0.0042 | $Ho_{7\_D}$ | Reject | 0.0014 | $Ho_{7\_A}$ | Reject |
| | Customers not served served | 0.0000 | $Ho_{8\_D}$ | Reject | 0.0000 | $Ho_{8\_A}$ | Reject |

Table 3 shows that all p-values for "waiting times" and "number of customers not served" in all four comparison tests are smaller than the chosen significance level (0.05). Thus, all sub-hypotheses for the comparison have to be rejected.

The statistical test results reveals the significant difference between the reactive behaviour results in Experiment 1 compared to Experiments 2, 3, 4 and 5 which consisted of mixed reactive and proactive behaviour modelling for both DES and combined DES.ABS models. Hence, $Ho_1$ and $Ho_2$ hypotheses are rejected.

From the comparison investigation of model result, new knowledge is obtained. Modelling mixed reactive and proactive behaviours in DES and combined DES/ABS models has a big impact on the simulation model output for this case study. In addition, DES and combined DES/ABS models produce similar simulation results for this case study.

## 6 Conclusions and Future Work

In this paper we have demonstrated the application of simulation to study the impact of human reactive and proactive behaviour service systems. In particular we were interested in finding out more about the benefits of implementing only reactive or mixed reactive and proactive behaviours. More precisely, our investigations have focused on answering the question: *Is it useful to model human proactive behaviour in service industry and which simulation techniques is the best choice for modelling such behaviour?*

Previously, we have dealt with the reactive behaviour modelling (Majid et al, 2009) and mixed reactive proactive behaviour modelling (Majid et al, 2010) in a first case study based in the retail sector. We found that modelling realistic proactive behaviours that habitually occur in the real system are worth modelling in the modelled situations as it has demonstrated a big impact to the overall system performance in DES and combined DES/ABS models.

In this paper, we have focused on modelling different level of proactive behaviour for our public service case study (Nottingham University International Office). Modelling the service-oriented system as realistically (proactive behaviour) as possible is found important because modelling such detail has a significant impact on the overall system performance – reducing customer waiting time and number of customers not served. Overall, DES and combined DES/ABS are found suitable for modelling the levels of proactive behaviour investigated in this case study. In the future we would like to start with the third case study, this time at the airport, to test if we can generalise our findings regarding the comparison of modelling different level of proactive behaviour in DES and combined DES/ABS techniques.

## Authors Biographies

MAZLINA ABDUL MAJID is a Senior Lecturer in the Faculty of Computer Systems and Software Engineering, Universiti Malaysia Pahang(UMP), Malaysia. Her research interest is in discrete event simulation and agent based simulation. Her email is <mazlina@ump.edu.my>.

PEER–OLAF SIEBERS is a Senior Research Fellow in the School of Computer Science, University of Nottingham, UK. The central theme in his work is the development of human behaviour models which can be used to better represent people and their behaviours in Operational Research type simulation models. His email is <pos@cs.nott.ac.uk>.

UWE AICKELIN is a Professor in the School of Computer Science, University of Nottingham, UK. His research interests include agent based simulation, heuristics optimisation, artificial immune system. His email is <uxa@cs.nott.ac.uk>.